\documentclass{ieeeaccess}
\usepackage{cite}
\usepackage{amsmath,amssymb,amsfonts}
\usepackage{algorithmic}
\usepackage{graphicx}
\usepackage{textcomp}
\usepackage{multirow}
\usepackage{bm}
\usepackage{url}
\usepackage{array}
\usepackage{makecell}

\makeatletter
\AtBeginDocument{\DeclareMathVersion{bold}
\SetSymbolFont{operators}{bold}{T1}{times}{b}{n}
\SetSymbolFont{NewLetters}{bold}{T1}{times}{b}{it}
\SetMathAlphabet{\mathrm}{bold}{T1}{times}{b}{n}
\SetMathAlphabet{\mathit}{bold}{T1}{times}{b}{it}
\SetMathAlphabet{\mathbf}{bold}{T1}{times}{b}{n}
\SetMathAlphabet{\mathtt}{bold}{OT1}{pcr}{b}{n}
\SetSymbolFont{symbols}{bold}{OMS}{cmsy}{b}{n}
\renewcommand\boldmath{\@nomath\boldmath\mathversion{bold}}}
\makeatother

\def\BibTeX{{\rm B\kern-.05em{\sc i\kern-.025em b}\kern-.08em
    T\kern-.1667em\lower.7ex\hbox{E}\kern-.125emX}}

\begin{document}
\history{Date of publication xxxx 00, 0000, date of current version xxxx 00, 0000.}
\doi{10.1109/ACCESS.2024.0429000}

\title{LMLCC-Net: A Semi-Supervised Deep Learning Model for Lung Nodule Malignancy Prediction from CT Scans using a Novel Hounsfield Unit-Based Intensity Filtering}
\author{
Tasnia~Binte~Mamun\authorrefmark{1},
Adhora~Madhuri\textsuperscript{‡}\authorrefmark{1},
Nusaiba~Sobir\textsuperscript{‡}\authorrefmark{1}, and
Taufiq~Hasan\authorrefmark{1}\IEEEmembership{Senior Member, IEEE}
}

\address[1]{mHealth Lab, Department of Biomedical Engineering, Bangladesh University of Engineering and Technology (BUET), Dhaka-1205, Bangladesh.}

\tfootnote{%
\textsuperscript{‡}These authors contributed equally.\\
Correspondence to: Taufiq Hasan (email: taufiq@bme.buet.ac.bd).
}

\markboth
{Mamun \headeretal: LMLCC-Net: A Semi-Supervised Deep Learning Model for Lung Nodule Malignancy Prediction}
{Mamun \headeretal: LMLCC-Net: A Semi-Supervised Deep Learning Model for Lung Nodule Malignancy Prediction}

\corresp{Corresponding author: Taufiq Hasan (e-mail: taufiq@bme.buet.ac.bd).}

\begin{abstract}
Lung cancer remains one of the leading causes of mortality worldwide. Early diagnosis of malignant pulmonary nodules in CT images can significantly reduce disease-related mortality and morbidity. This study aims to develop a novel deep learning framework that leverages the unique intensity characteristics of nodules, measured in Hounsfield Units, for improved classification of benign and malignant pulmonary nodules. In this work, we propose LMLCC-Net, a novel deep learning framework for classifying nodules from CT scan images using a three-dimensional convolutional neural network, incorporating Hounsfield Unit based intensity filtering. Benign and malignant nodules exhibit significant differences in their Hounsfield Unit based intensity profiles, which have not been sufficiently explored in the existing literatures. Our method considers both the intensity pattern and the texture for malignancy prediction. LMLCC-Net extracts features from multiple branches, each using a separate, learnable Hounsfield Unit based intensity filtering stage. Various combinations of branches and learnable filter ranges were explored to identify the best-performing model. In addi tion, we propose a semi-supervised learning scheme for labeling ambiguous cases and develop a lightweight model to efficiently classify the nodules. The proposed LMLCC-Net was evaluated using the LUNA16 dataset. Our proposed method achieves a classification accuracy of 91.96\%, a sensitivity of 92.94\%, and an area under the curve of 94.07\%, showing improved performance compared to existing methods. Our proposed LMLCC-Net framework, which effectively combines Hounsfield Unit-based intensity filtering and texture analysis, significantly improves the classification performance of pulmonary nodules in CT images. By utilizing a multi branch architecture and semi-supervised learning, the model enhances its ability to handle ambiguous cases while maintaining computational efficiency through a lightweight design.
\end{abstract}


\titlepgskip=-21pt

\maketitle

\section{Introduction}

\label{sec:introduction}
\par{L}{ung} cancer is the leading cause of cancer deaths worldwide. It was the most frequently diagnosed cancer in 2022, responsible for almost 2.5 million new cases, or one in eight cancers worldwide (12.4\% of all cancers globally) \cite{bray2024global}. Early detection is crucial to improving the survival rates of lung cancer patients, as it allows for effective and timely treatment interventions \cite{parkin2000lung}. Low-dose computed tomography (CT) screening has emerged as a powerful tool for the early detection of lung cancer, enabling the identification of lung nodules that may indicate malignancy \cite{rubin2015lung}. However, the implementation of annual low-dose CT screenings substantially increases radiologists' workload, particularly in developing countries where there is a pronounced shortage of expert and qualified radiologists \cite{lancaster2022low}. Another challenge is that lung nodules, which can be solid, partially solid, or ground glass, each with varying cancer probabilities, exhibit extraordinary spatial and temporal heterogeneity at multiple levels, including genes, proteins, cells, microenvironment, tissues, and organs \cite{loverdos2019lung}, leading to missed diagnoses and reporting errors. This underscores the critical need for efficient, accurate and automated methods to assist radiologists in the detection and prediction of lung nodule malignancy. 

\begin{figure*}[!t]
    \centering
    \begin{minipage}[t]{0.05\linewidth}
        \centering
        \vspace{-70pt} 
        \hspace{-3pt} 
    \end{minipage}
    \begin{minipage}[t]{0.9\linewidth}  % This controls the image width
        \centering
        \includegraphics[width=\linewidth]{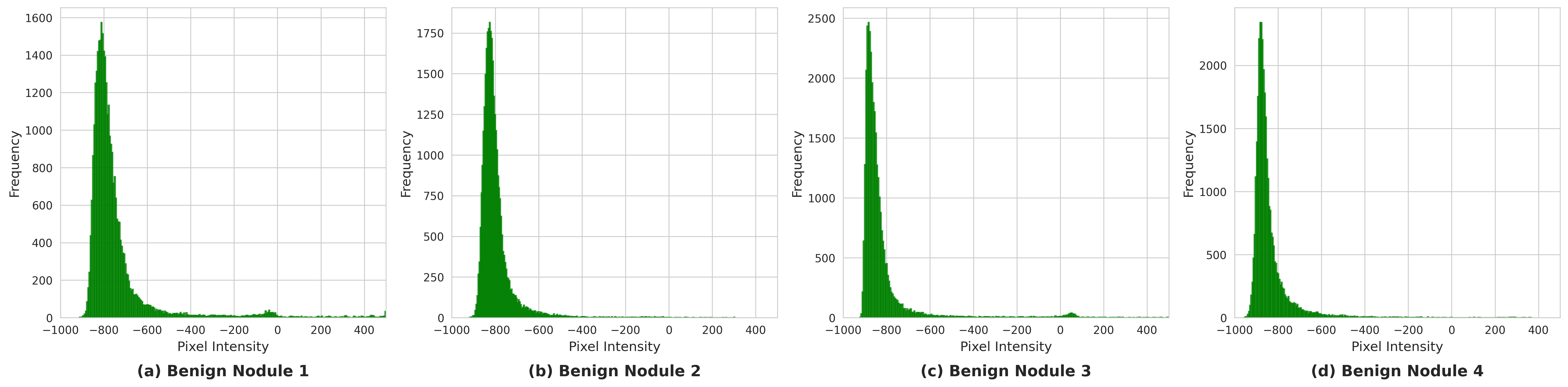}
    \end{minipage}   

    \vspace{10pt} 

    \begin{minipage}[t]{0.05\linewidth}
        \centering
        \vspace{-70pt} 
        \hspace{-3pt} 
    \end{minipage}
    \begin{minipage}[t]{0.9\linewidth}  % This controls the image width
        \centering
        \includegraphics[width=\linewidth]{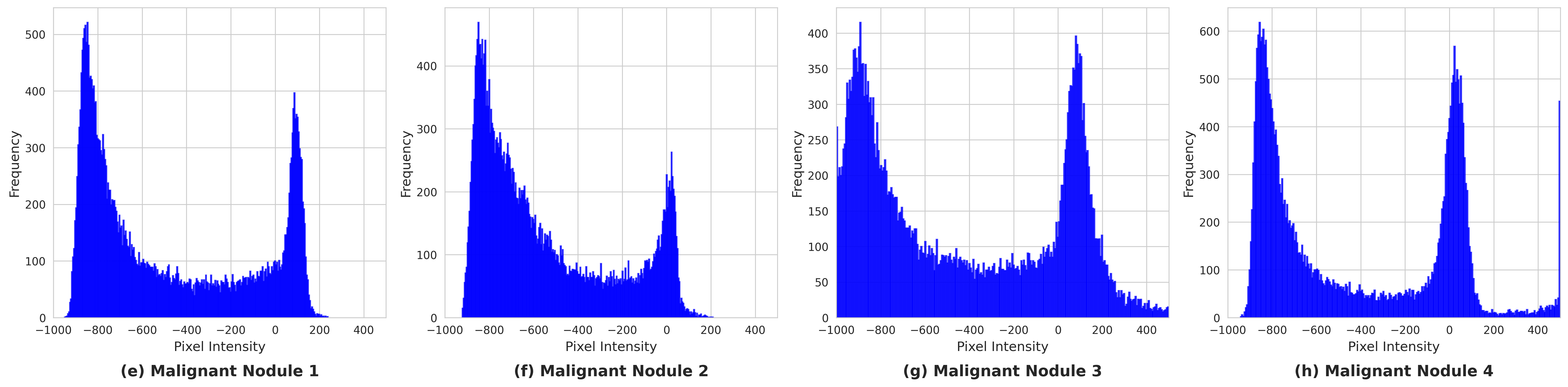}
    \end{minipage}

    \caption{Intensity range corresponding to the Hounsfield Unit (HU) distribution: (a-d) benign nodules and (e-h) malignant nodules.}
    \label{f1}
\end{figure*}

Most of the previous studies on convolutional neural network (CNN) based lung nodule analysis methods \cite{guo20233d, kumar2021classification, dai2018incorporating, kuang2020unsupervised, joshi2021lung, causey2018highly, apostolopoulos2021classification, qin2021relationship, zhang2022re, tang2021classification} in this area have used the publicly available Lung Image Database Consortium image collection (LIDC-IDRI) dataset \cite{messay2015segmentation}. These studies calculated the mean malignancy level (MeanML) for each nodule. However, this labeling method has limitations in accuracy. A significant amount of data with potential labeling inaccuracies can adversely affect model training.
In addition, no prior study has explored the distribution patterns of HU within lung nodules or considered focusing on distinct HU ranges as a feature for classification. Benign tumors typically exhibit smooth, regular borders, while malignant tumors often present with irregular borders \cite{tumorlist, Benigntumors, patel2020benign}. Additionally, malignant tumors have been shown to possess higher values of vascularity index and maximal shear velocity, indicating that malignant lesions tend to be stiffer compared to benign or intermediate tumors \cite{ohshika2021distinction}. These characteristics influence the HU distribution within nodules across different ranges, making it a valuable feature for distinguishing between benign and malignant tumors. Example distributions of HU intensity profiles in benign and malignant tumors are represented in Fig.~\ref{f1}.

\begin{figure}[!h] 
  \includegraphics[width=\linewidth]{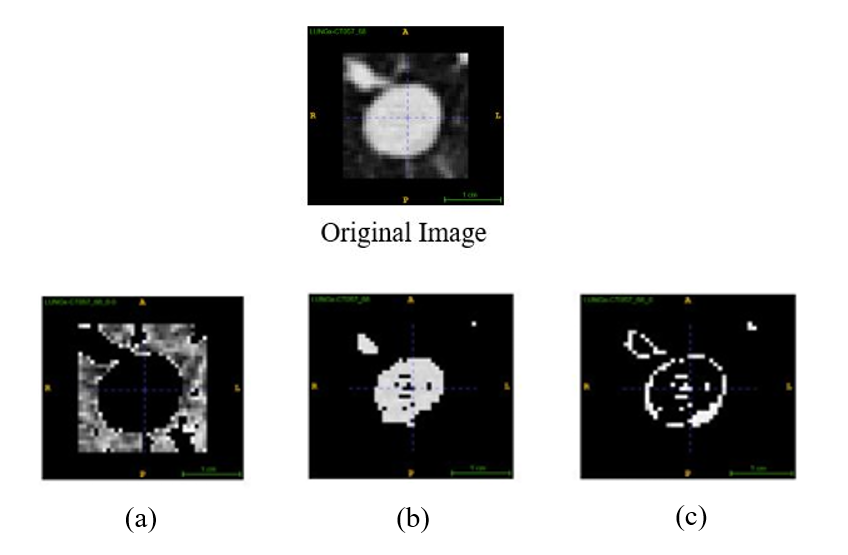}
  \caption{Visualization of HU range segmentation for a representative nodule. (a) HU range divided between 0–0.3, highlighting low-intensity regions; (b) 0.3–0.7, capturing mid-range densities; and (c) 0.8–1.0, emphasizing high-density structures. This division helps isolate subtle tissue-specific variations that support malignancy prediction.}
  \label{HU}

\end{figure}

Furthermore, to better visualize how the Hounsfield Unit-based segmentation captures intensity-specific regions within a nodule, representative images were generated by segmenting the HU ranges into multiple sub-intervals. As shown in Fig. \ref{HU}, this process isolates intensity-dependent structural variations that correspond to different tissue densities, allowing the model to focus on diagnostically relevant sub-regions such as soft tissue, fibrotic cores, and calcified components.

The main contributions of our study can be summarized as follows.
\vspace{0.15cm}

1)	We develop a novel approach to labeling the LUNA16 dataset, implementing a semi-supervised method specifically designed to handle ambiguous cases. This strategy enhances the accuracy of the labels, resulting in more reliable data for model training.

2) We propose the Learnable Multi-branched Lung Cancer Classification Network (LMLCC-Net), which processes CT volumes through multiple HU-based branches to extract complementary texture and density information for malignancy prediction. We intoduce a Learnable Dynamic Range Layer within each branch that automatically optimizes HU intensity boundaries during training, enabling adaptive feature extraction and enhancing generalization across heterogeneous nodules.

3) We conduct extensive validation across multiple architectural configurations including different multibranch setups, learnable versus fixed HU ranges, and constant versus random initialization strategies to confirm the stability and generalization of the proposed framework. Grad-CAM–based interpretability analysis further verifies that the network consistently attends to diagnostically relevant nodule regions. This systematic evaluation ensures the model’s robustness and reliability, making it well-suited for deployment in real-world digital imaging and clinical screening environments.

\vspace{0.15cm}

\vspace{0.15cm}
The remainder of this paper is structured as follows: Section \ref{sec:Related Works} includes a brief overview of the related works. Section \ref{sec:Dataset} describes the dataset. Section \ref{sec:method} provides a detailed description of the proposed model. Section \ref{sec:Experimental Setup} features a comprehensive overview of the training parameters and dataset splitting. Section \ref{sec:results and analysis} presents our results and analysis. Finally, Section \ref{sec:conclusion} covers related issues and concludes with our findings.

\section{Related Works}
\label{sec:Related Works}
Recent advancements in conventional computer-aided diagnosis (CAD) systems have shown significant promise in automating the detection and classification of pulmonary nodules.
Conventional computer-aided diagnosis (CAD) systems rely on extracting handcrafted features from lung nodules to build radiomics models for automatic classification \cite{ojala2002multiresolution, chen2018radiomic,thawani2018radiomics}. In this approach, various feature sets have been utilized to train classification algorithms like random forests, support vector machines (SVM), and linear discriminant analysis (LDA). While these methods can yield promising results for well-defined nodules, they have a significant drawback, which is their limited capacity to handle the diverse sizes, contexts, and shapes of different nodules. Publicly available datasets and competitions, such as LUNA16, Kaggle DSB 2017, and LIDC-IDRI, have facilitated the rapid development of advanced computerized systems for this purpose, primarily using deep learning-based algorithms. CNNs have proven to be a strong alternative method for classifying nodules. Wang et al. \cite{wang2023diagnostic} developed MResNet, which is based on ResNet and PPM. MResNet extracts general features from CT images for accurate classification of lung nodules. Using limited chest CT data, Xie et al. \cite{xie2018knowledge} proposed a multi-view knowledge-based collaborative (MV-KBC) deep model to distinguish malignant from benign nodules. The model decomposes 3D nodules into nine views, with each view processed by a knowledge-based collaborative (KBC) sub-model using pre-trained ResNet-50 networks. Asuntha et al. \cite{asuntha2020deep} developed a method that extracts four types of features: texture, geometric, volumetric, and intensity, employing techniques like wavelet transform, Local Binary Pattern (LBP), Scale Invariant Feature Transform (SIFT), and Zernike Moment. After extracting features, the Fuzzy Particle Swarm Optimization (FPSO) algorithm is applied to select the best feature. In order to make the results interpretable, Shen et al. \cite{shen2019interpretable} provided a novel interpretable deep hierarchical semantic convolutional neural network (HSCNN) that has two levels, a low-level and a high-level schematic. Combining both of them produces an interpretable output. There is a necessity for careful analysis of the suspected nodules. For this purpose, the model proposed by Jifara et al. \cite{sori2021dfd} has two branches: a residual learning denoising model(DR-Net) employed to remove noise and a two-path convolutional neural network which takes the denoised image output from DR-Net and detects lung cancer. 

A two-dimensional (2D) convolutional neural network can only capture features from individual slices and cannot fully utilize the three-dimensional (3D) information present in CT images. Xiuyuan et al. \cite{xu2020mscs} utilized MSCSDeepLN that analyzes lung nodule malignancy. Lung nodules’ malignancy is assessed by combining three trained light models. Each light model uses three-dimensional convolutional neural networks (CNNs) as its foundation to extract lung nodule features from CT images while maintaining lung nodule spatial heterogeneity. The sub-networks can learn the multi-level contextual features and preserve variety by using multi-scale input cuts from CT images. Iftikhar et al. \cite{naseer2023lung} proposed a lung cancer classification method involving a three-phase process utilizing modified U-Net and AlexNet architectures. In the first phase, using predicted masks, a modified U-Net segments lung lobes from CT slices. The second phase uses another modified U-Net to extract candidate nodules based on these masks and labels. In the third phase, a modified AlexNet combined with a support vector machine (SVM) classifies the candidate nodules as cancerous or non-cancerous. Li et al. \cite{li2019predicting} integrated handcrafted features (HF) with those learned from the output layer of a 3D deep CNN. In this approach, the CNN-derived features, along with 29 handcrafted features, were used as input to an SVM, paired with the sequential forward feature selection (SFS) method, to optimize the feature subset and build the classifiers. Zhang et al. \cite{zhang2019lung} adapted VGG19, VGG16, MobileNet, DenseNet121, ResNet50, Xception, NASNetLarge, and NASNetMobile into 3D-CNN models and tested them on the LIDC-IDRI lung CT dataset, . The experimental results indicated that Xception and DenseNet121 outperformed the others in lung nodule diagnosis, achieving superior sensitivity, specificity, accuracy, area under the curve (AUC), and precision.

\section{Dataset}
\label{sec:Dataset}
In this study, we use the LUNA16 dataset (Lung Nodule Analysis 2016)\cite{luna}, which is a subset of the LIDC-IDRI dataset \cite{lidc}. LIDC-IDRI consists of thoracic CT scans along with annotations provided by multiple radiologists. It contains 1018 low-dose CT images. LIDC-IDRI contains all the relevant information about the CT scan, including the case name, the nodule centroid coordinates, diameter, the nodule volume, nodule texture, and the marking results from all four radiologists. The LUNA16 dataset removes the CT scans with a slice thickness greater than 2.5 mm. Thicker slices can reduce image resolution and make identifying and measuring small nodules challenging. Nodules that are 3 mm or larger and accepted by at least three of four radiologists are included in the reference standard. Thus, nodules smaller than 3 mm or annotated by only one or two radiologists are excluded. Finally, 888 low-dose CT images are generated. The LUNA16 dataset did not initially include malignancy labels. For our preparation, we updated it with the radiologists’ malignancy markings corresponding to patient IDs from the metadata CSV file provided with the LIDC-IDRI dataset. In the LIDC-IDRI dataset, nodules are marked from 1-5, 1 being benign and 5 being the highest malignancy level. According to the malignancy score, the annotated nodule with a score less than three is benign, a score equal to three is uncertain, and a score greater than three is malignant. 

\section{Method}
\label{sec:method}
\subsection{Dataset Preparation}
Unlike existing methods, we do not average the radiologists’ annotations to determine the labels for CT scans, as it often results in ambiguous or suboptimal labeling decisions. We establish specific criteria based on the consensus or majority opinion of the radiologists. Our decision criteria for a nodule to be considered malignant are as follows:
\begin{itemize}
    \item When a nodule is annotated by four radiologists: At least three radiologists must assign a rating greater than three, or one radiologist assigns a rating of three, while at least two others assign ratings greater than three.
    \item When a nodule is annotated by three radiologists: At least two radiologists must assign a rating greater than three
    \item When a nodule is annotated by two radiologists: Both radiologists must assign ratings greater than three.
\end{itemize}
The same criteria are used to classify a nodule as benign; the only difference is that the annotation of less than three instead of over three will be considered in this case. The labeling criteria are further explained in Table~\ref{tab:criteria}.

\begin{table}[ht]
\scriptsize
\centering
\renewcommand{\arraystretch}{1.5}
\caption{\textbf{Criteria for a nodule to be considered as malignant or benign.}}
\label{tab:criteria}
\begin{tabular}{>{\centering\arraybackslash}p{14mm}|p{14mm}|p{14mm}|p{14mm}|c} 
\hline \begin{tabular}{c} 

\textbf{No. of}\\
\textbf{radiologists'}\\
\textbf{annotations}
\end{tabular} & 
\begin{tabular}{c} 
\textbf{No. of}\\
\textbf{ratings$>3$}
\end{tabular} & 
\begin{tabular}{c} 
\textbf{No. of} \\
\textbf{ratings$=3$}
\end{tabular} & 
\begin{tabular}{c} 
\textbf{No. of} \\
\textbf{ratings$<3$}
\end{tabular} & 
\textbf{Label} \\

\hline 

\multirow{8}{*} {\makecell{\centering 4}} & \centering 4 & \centering 0 & \centering 0 & \multirow{4}{*}{1} \\  
& \centering 3 & \centering 1 & \centering 0 & \\  
& \centering 3 & \centering 0 & \centering 1 & \\  
& \centering 2 & \centering 1 & \centering 1 & \\\cline{2-5}
& \centering 0 & \centering 0 & \centering 4 & \multirow{4}{*}{0} \\  
& \centering 0 & \centering 1 & \centering 3 & \\  
& \centering 1 & \centering 0 & \centering 3 & \\  
& \centering 1 & \centering 1 & \centering 2 & \\  

\hline 
\multirow{6}{*}{\makecell{3}} & \centering 3 & \centering 0 & \centering 0 & \multirow{3}{*}{1} \\  
& \centering 2 & \centering 1 & \centering 0 & \\  
& \centering 2 & \centering 0 & \centering 1 & \\\cline{2-5} 
& \centering 0 & \centering 0 & \centering 3 & \multirow{3}{*}{0} \\  
& \centering 0 & \centering 1 & \centering 2 & \\  
& \centering 1 & \centering 0 & \centering 2 & \\  

\hline 
\multirow{2}{*}{\makecell{2}} & \centering 2 & \centering 0 & \centering 0 & 1 \\\cline{2-5}
& \centering 0 &\centering 0 & \centering 2 & 0 \\  
\hline
\end{tabular}
\end{table}

\begin{figure*}[!h]
    \centering
    \begin{minipage}[t]{0.05\linewidth}
        \centering
        \textbf{(a)}
        %\vspace{-10pt} 
        %\hspace{-3pt} 
        
    \end{minipage}
    \begin{minipage}[t]{0.95\linewidth}
        \centering
        \includegraphics[width=\linewidth]{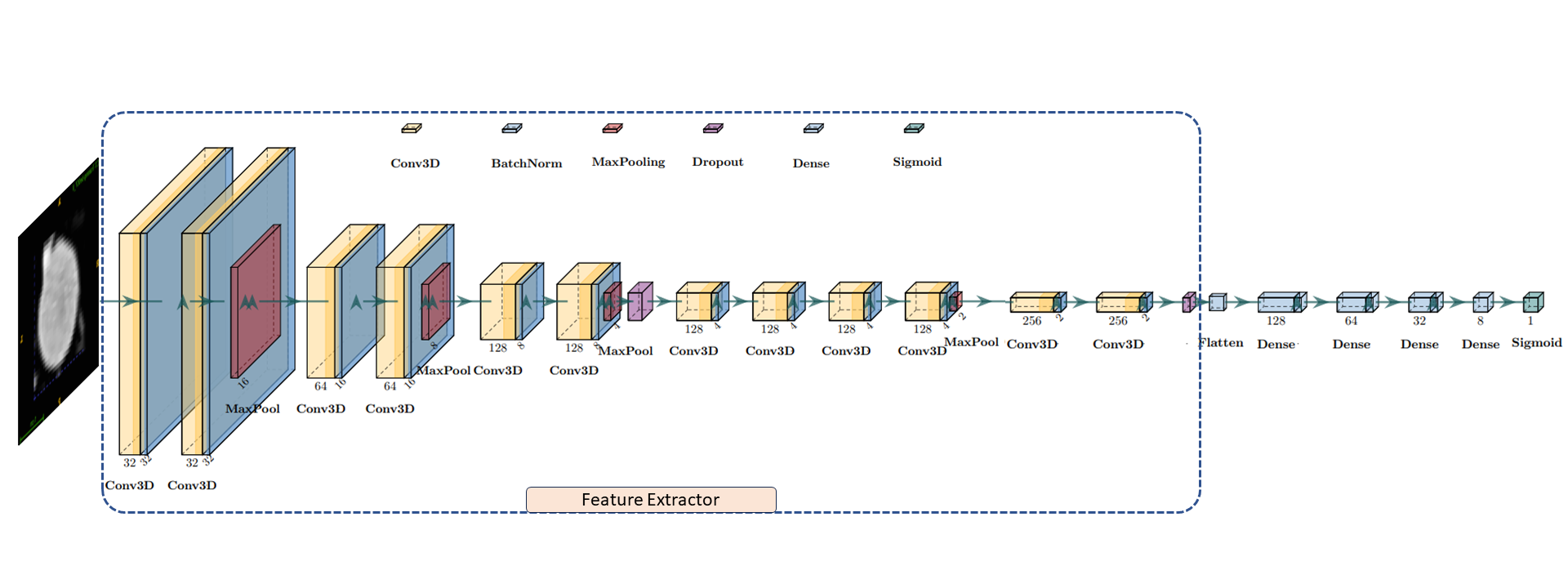}
    \end{minipage}   

    \vspace{5pt}  

    \begin{minipage}[t]{0.05\linewidth}
        \centering
        \vspace{-10pt} 
        \hspace{-3pt} 
        \textbf{(b)}
    \end{minipage}
    \begin{minipage}[t]{0.95\linewidth}
        \centering
        \includegraphics[width=\linewidth]{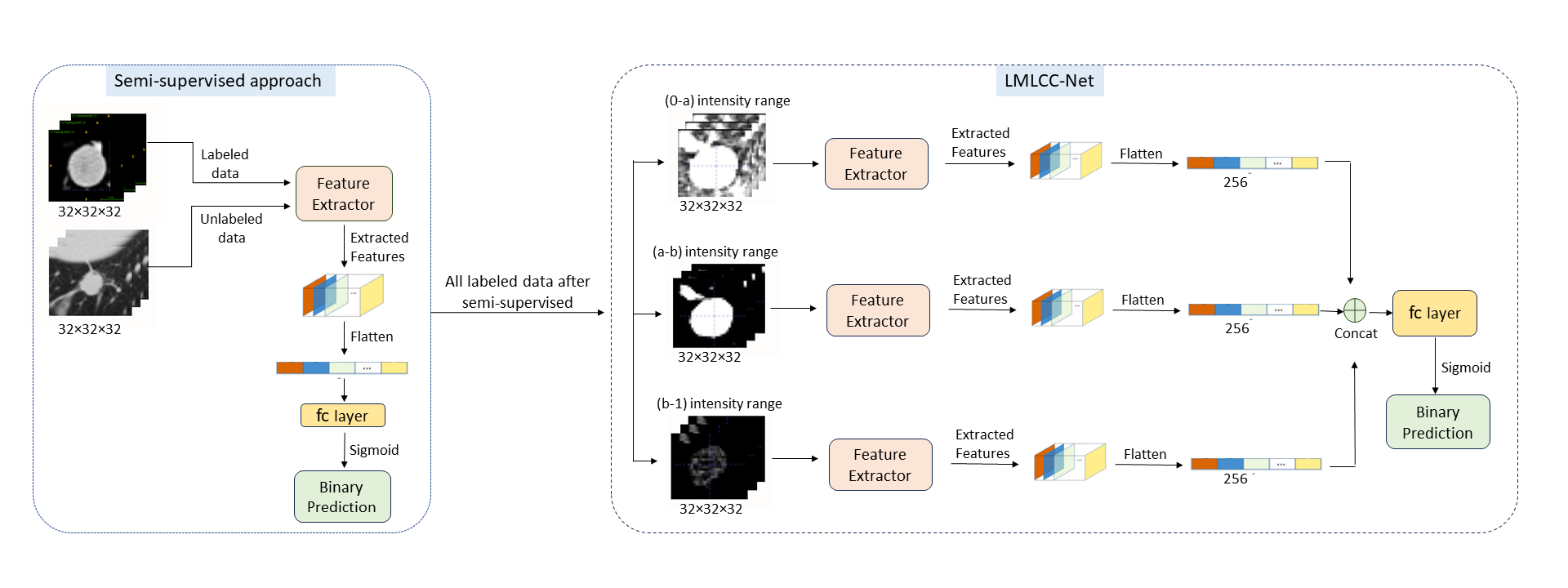}
    \end{minipage}

    \caption{(a) Backbone 3D CNN model  (b) Pipeline of our proposed framework, LMLCC-Net}
    \label{f2} 
\end{figure*}

We also consider an exclusion criterion. If none of the specified conditions for classifying a nodule as either malignant or benign are met, we label the nodule as ambiguous. We refine our dataset to include 558 nodules with clear and necessary malignancy information by applying these criteria, and the remaining 612 nodules with ambiguous malignancy annotations are retained for semi-supervised learning, allowing us to determine their labels with greater confidence. This novel semi-supervised technique successfully reduces the ambiguity of the dataset and increases the number of labeled instances.

\subsection{Preprocessing} 
Typically, radiodensities of various parts of a CT scan are different\cite{HU}. Thus, the pixel intensity range is rescaled between -1000 and 500 to leave the lung nodule as the only segment. Next, we normalize the intensity between 0 to 1. Most nodules have a diameter of 4–8 mm and an average diameter of 8.3 mm \cite{guo20233d}. We first extract three cubic patches of dimensions $32 \times 32 \times 32$, $48 \times 48 \times 48$, and $64 \times 64 \times 64$ voxels centered at each candidate location. Since chest CT scans have a variable spatial resolution, we resample to voxel spacing $0.7\times0.7\times1$ mm$^3$. Each scan is rotated in 7 angles, each spaced 45 degrees apart that is 45°, 90°, 135°, 180°, 225°, 270°, 315°.

%\subsection{Proposed Learnable Multi-branched Lung Cancer Classification Model (LMLCC-Net)}
\subsection{Proposed LMLCC-Net}
The architecture of the proposed backbone 3D CNN model is shown in Fig.~\ref{f2}. It uses binary cross-entropy loss, and the CNNs use ReLU activation and batch normalization for hierarchical feature extraction after each convolutional layer during training. 

% \begin{figure}[!h]
%     \centering
%     \begin{minipage}[t]{0.5\linewidth}
%         \centering
%         \vspace{-10pt} 
%         \hspace{-3pt} 
%         \textbf{(a)}
%     \end{minipage}
%     \begin{minipage}[t]{1\linewidth}
%         \centering
%         \includegraphics[width=\linewidth]{model.png}

%     \end{minipage}   

%     \vspace{5pt}  

%     \begin{minipage}[t]{0.5\linewidth}
%         \centering
%         \vspace{-10pt} 
%         \hspace{-3pt} 
%         \textbf{(b)}
%     \end{minipage}
%     \begin{minipage}[t]{1\linewidth}
%         \centering
%         \includegraphics[width=\linewidth]{modelall.png}
 
%     \end{minipage}

%     \caption{(a) Backbone 3D CNN model  (b) Pipeline of our proposed framework, LMLCC-Net }
%     \label{f2} 
% \end{figure}

The images are first fed into the feature extractor, which consists of twelve 3-D convolution layers with filters of size 3 × 3 × 3. It also employs four MaxPooling and two Dropout layers to reduce dimensionality and prevent overfitting, respectively. After the feature extractor, the model flattens the extracted features and sends them through five fully connected (dense) layers. The last fully connected layer is followed by a sigmoid activation, which gives the final binary output.

The proposed LMLCC-Net is a custom 3D convolutional neural network (CNN) designed based on a backbone 3D CNN model for efficient feature extraction from volumetric data. Traditional CNNs often struggle to adapt to the wide range of intensities present in 3D data, leading to suboptimal performance. To overcome this limitation, we introduced a novel Learnable Dynamic Range Layer that dynamically adjusts the input range, enhancing the model's ability to capture relevant features across different intensity levels. This custom layer splits the input into multiple branches based on the learned threshold, enabling the model to capture variations in intensity within the data better. Following the convolutional blocks, the feature maps from all branches are flattened and concatenated, creating a unified representation of the input data. This representation is passed through several fully connected layers, culminating in a sigmoid activation for binary classification. 
During branching, the branched input goes through individual models. The model extracts information from each of the branches and concatenates them. As each branch focuses on a specific HU-intensity range, the data extraction is more effective, and the model can classify the tumor more efficiently.

\section{Experimental Setup}
\label{sec:Experimental Setup}
To train the proposed network, we adopt the binary cross-entropy loss. Binary cross-entropy compares each of the predicted probabilities to the actual class output, which can be either 0 or 1.

The binary cross-entropy (log) loss is defined as:

\begin{equation}
\text{Loss} = - \frac{1}{N} \sum_{i=1}^{N} \Big[ y_i \log \hat{y}_i + (1 - y_i) \log (1 - \hat{y}_i) \Big]
\end{equation}

where $y_i$ denotes the actual value and $\hat{y}_i$ is the predicted probability output from the neural network.

We use Adam optimizer to train the model for 200 epochs with a batch size of 68. The minimum learning rate is $1\times10^{-6}$. The train and validation set contains 80\% of labeled data of LUNA16. Among 80\% of labeled data, the train set contains 80\%, and the validation set contains 20\%. The test set contains 20\% of labeled LUNA16. We divided the data into test, train, and validation sets based on nodule ID.

\begin{figure}[!h] 
  \includegraphics[width=3.5 in,height=2.5in,clip,keepaspectratio]{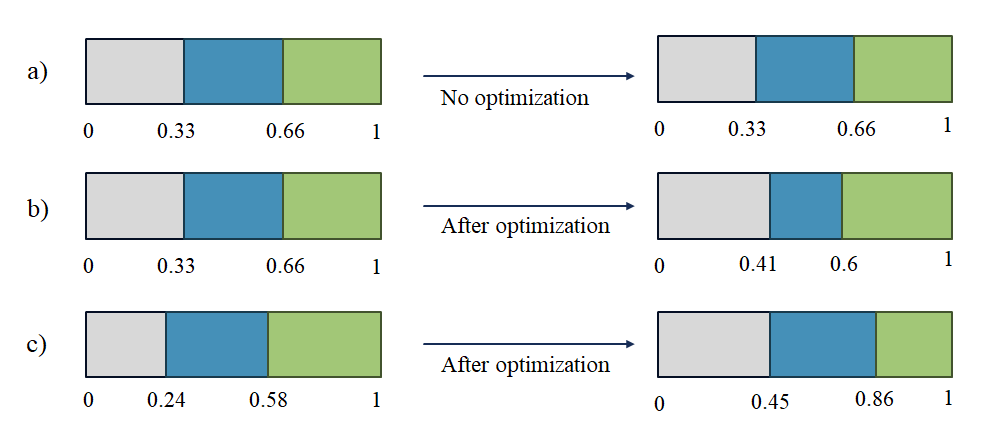}
  \caption{a) Non-learnable- The Input in which the intensity ranges should be divided will be mentioned to the classifier. b) Learnable (Constant initialization)- The number of branches of intensity division and ranges will be mentioned. c) Learnable (Random initialization)- Only the number of branches of intensity division will be mentioned.}
  \label{f3}
\end{figure}

Data that we eliminated due to inadequate annotation are considered unlabeled data in semi-supervised learning. We continue the training until the model labels most of the data with more than 90\% confidence. Validation and test sets are fixed in these trainings. Next, we try various intensity range filters on 3D data in multiple ways, to analyze the effect of model performance on different parameter setting which is shown in Fig.~\ref{f3}.

\begin{figure*}[!h]
    \centering
    % --- (a)
    \begin{minipage}[t]{0.05\linewidth}
        \centering
        \vspace{-10pt}
        \hspace{-3pt}
        \textbf{(a)}
    \end{minipage}%
    \begin{minipage}[t]{0.9\linewidth}
        \centering
        \includegraphics[width=\linewidth]{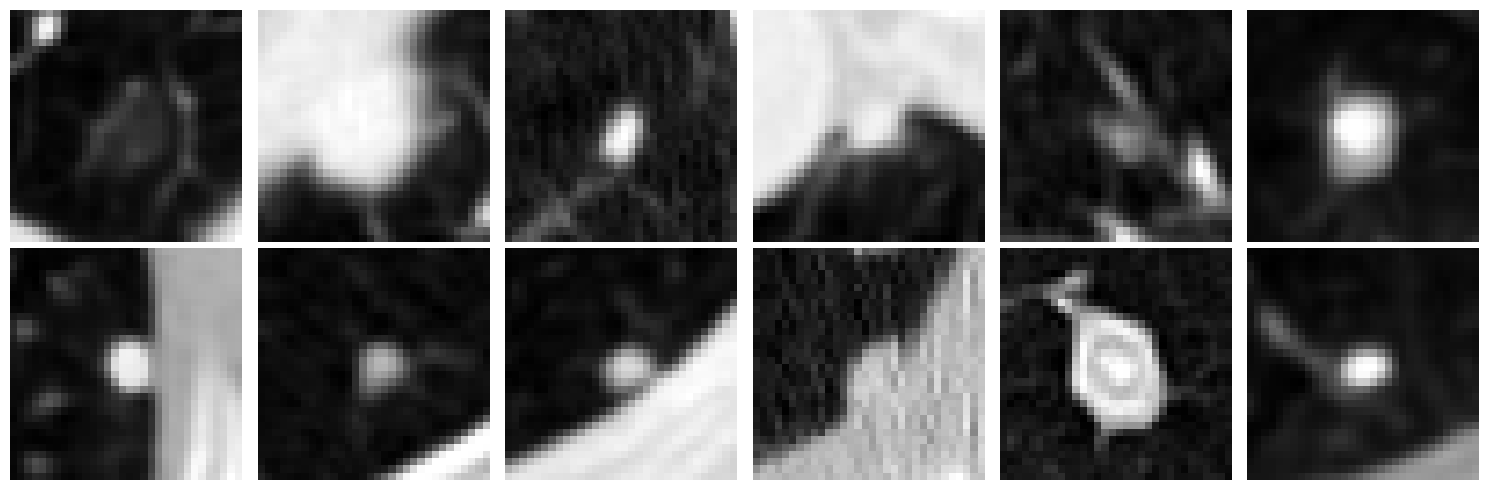}
    \end{minipage}

    \vspace{10pt} % space between (a) and (b)

    % --- (b)
    \begin{minipage}[t]{0.05\linewidth}
        \centering
        \vspace{-10pt}
        \hspace{-3pt}
        \textbf{(b)}
    \end{minipage}%
    \begin{minipage}[t]{0.9\linewidth}
        \centering
        \includegraphics[width=\linewidth]{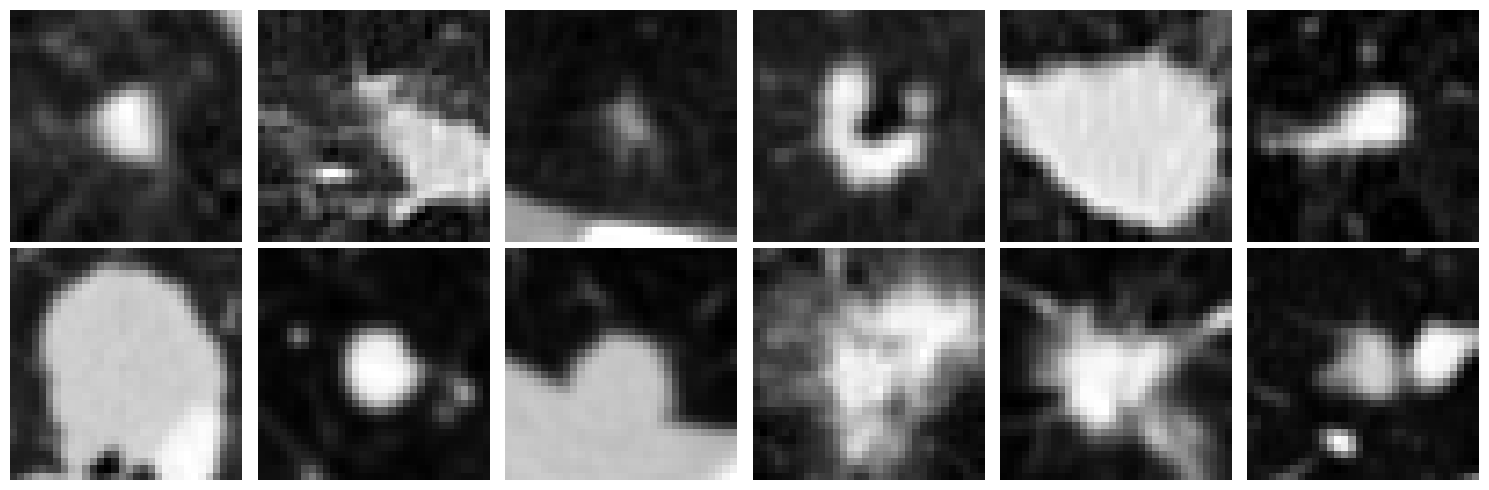}
    \end{minipage}
    \caption{Examples of correctly classified nodules: (a) benign nodules and (b) malignant nodules.}
    \label{f4}
\end{figure*}
Introducing a learnable HU-intensity filter is a key innovation of our work. We examined the performance of the learning parameter with both random initialization and constant initialization. We observed that the model provides improved results when we make the interval learnable instead of using a non-learner model. The input at various intensity ranges is created by making a custom function. In this function, we specify the number of branches that are needed. The function randomly or constantly initializes in that many intervals within a constraint and, upon training, optimizes that HU-filter interval.

\section{Results and Analysis}
\label{sec:results and analysis}
\subsection{Evaluation Metrics}

Evaluation metrics are critical tools for assessing the performance of models. Classifiers divide classified samples into four categories: TP (True Positive) defines positive samples correctly recognized by the model. FP (False Positive) represents the number of negative samples recognized as positive by the model. TN (True Negative) represents the number of negative samples correctly identified by the model. FN (False Negative) represents the number of positive samples identified as negative by the model.

Using these definitions, several evaluation indicators are used to measure the model’s performance: accuracy, precision, sensitivity, specificity, and the area under the ROC curve. Definitions of these metrics are provided below:

\subsection*{Accuracy}
The overall classification accuracy ($A_{CC}$) is calculated as:

\begin{equation}
A_{CC} = \frac{TP + TN}{TP + TN + FP + FN}
\end{equation}

\subsection*{Precision}
Precision ($P_{RE}$), also known as the positive predictive value, is given by:
\begin{equation}
P_{RE} = \frac{TP}{TP + FP}
\end{equation}

\subsection*{Sensitivity}
Sensitivity ($S_{EN}$) measures the proportion of actual positives correctly identified:
\begin{equation}
S_{EN} = \frac{TP}{TP + FN}
\end{equation}

\subsection*{Specificity}
Specificity ($S_{PE}$) evaluates the proportion of actual negatives correctly identified:
\begin{equation}
S_{PE} = \frac{TN}{TN + FP}
\end{equation}

 The ROC curve or Receiver Operating Characteristic curve graphically represents and evaluates the performance of a binary classification model by illustrating the trade-off between the model's sensitivity and specificity in various settings. The Area Under the ROC Curve (AUC) represents the model's ability to discriminate between positive and negative classes. A higher AUC value indicates better overall performance of the classifier. An AUC of 0.5 suggests no distinguishing power, while an AUC of 1.0 indicates perfect classification.

\subsection{Experimental Results}
The majority of the research work regarding lung cancer detection from nodules in CT scans leverages the textural and morphological differences between benign and \mbox{malignant} nodules. For instance, benign nodules normally exhibit a smoother surface, higher sphericity, smoother edges, and minimal lobulation. On the contrary, malignant nodules are often characterized by uneven surfaces, irregular shapes, and the presence of spikes. The backbone 3D CNN correctly identifies the benign and malignant nodules depending on morphological features. Fig.~\ref{f4} presents 2D slices of some correctly \mbox{classified} nodules using backbone 3D CNN for qualitative assessment. As depicted, the proposed model accurately classifies the nodules, even when there are no apparent morphological or size differences, demonstrating backbone 3D CNN’s ability to distinguish between benign and malignant nodules using comprehensive feature analysis.

% \begin{figure}[!h]
%     \centering
%     \begin{minipage}[t]{0.05\linewidth}
%         \centering
%         \vspace{-10pt} 
%         \hspace{-3pt} 
%         \textbf{(a)}
%     \end{minipage}
%     \begin{minipage}[t]{1\linewidth}
%         \centering
%         \includegraphics[width=\linewidth]{CorrectBenignNodule.png}
%     \end{minipage}   
% \end{figure}
% \begin{figure}[!h]
%     \centering
%     \begin{minipage}[t]{0.05\linewidth}
%         \centering
%         \vspace{-10pt} 
%         \hspace{-3pt} 
%         \textbf{(b)}
%     \end{minipage}
%     \begin{minipage}[t]{1\linewidth}
%         \centering
%         \includegraphics[width=\linewidth]{CorrectMalignantNodule.png}
%     \end{minipage}
%     \caption{Examples of correctly classified nodules: (a) benign nodules and (b) malignant nodules.}
%     \label{f4}
% \end{figure}

\subsubsection{ Baseline model (Fully Supervised)}
%\vspace{0.5em}
To assess the performance of the proposed fully supervised baseline model, which contains a  3D CNN backbone, we conduct an evaluation of the LUNA16 dataset, comparing our results with several previously well-performing models on the same dataset. Table~\ref{tab:diff_model_compare} summarizes the comparative results. We train the models using the updated version of LUNA16, which is a confidently labeled dataset consisting of 558 samples. Table~\ref{tab:diff_model_compare} show that Baseline 3D CNN has improved $A_{CC}$, $P_{RE}$, $S_{EN}$ than other methods. Thus, the results show that our proposed baseline model outperforms other leading models in classifying benign and malignant nodules.

\begin{table}[ht]
\centering
\small 
\caption{\textbf{Performance Comparison of Models on the LUNA16 Dataset for Patch Size $32 \times 32 \times 32$} \label{tab:diff_model_compare}} 
\setlength{\tabcolsep}{6pt}  
\begin{tabular}{p{90pt} |p{20pt}| p{20pt}| p{20pt}| p{20pt}}
\hline
\textbf{Model} & \textbf{\boldmath$\%A_{CC}$} & \textbf{\boldmath$\%P_{RE}$} & \textbf{\boldmath$\%S_{EN}$} & \textbf{\%AUC} \\
\hline
Volumetric 3D CNN \cite{zunair2020uniformizing}  & 82.14  & 82.85  & 82.14  & 89.37 \\
Multi-View CNN \cite{setio2016pulmonary} & 83.00 & 84.00 & 84.00 & 90.00 \\
\textbf{Baseline 3D CNN} & \textbf{87.50} & \textbf{89.00} & \textbf{88.00} & {93.63} \\

DenseNet3D & 86.61 & 87.50 & 86.75 & 93.40 \\
ConvNeXtTiny & 81.25 & 81.25 & 81.26 & 89.28 \\
EfficientNet & 81.25 & 82.44 & 81.42 & 91.74 \\
MobileNet V2 & 85.71 & 86.96 & 85.55 & \textbf{94.99} \\
VGG-19 & 83.93 & 85.47 & 84.15 & 93.52 \\
\hline
\end{tabular}
\end{table}

\begin{figure}[!h] 
  \includegraphics[width=3.55 in,height=3 in,clip,keepaspectratio]{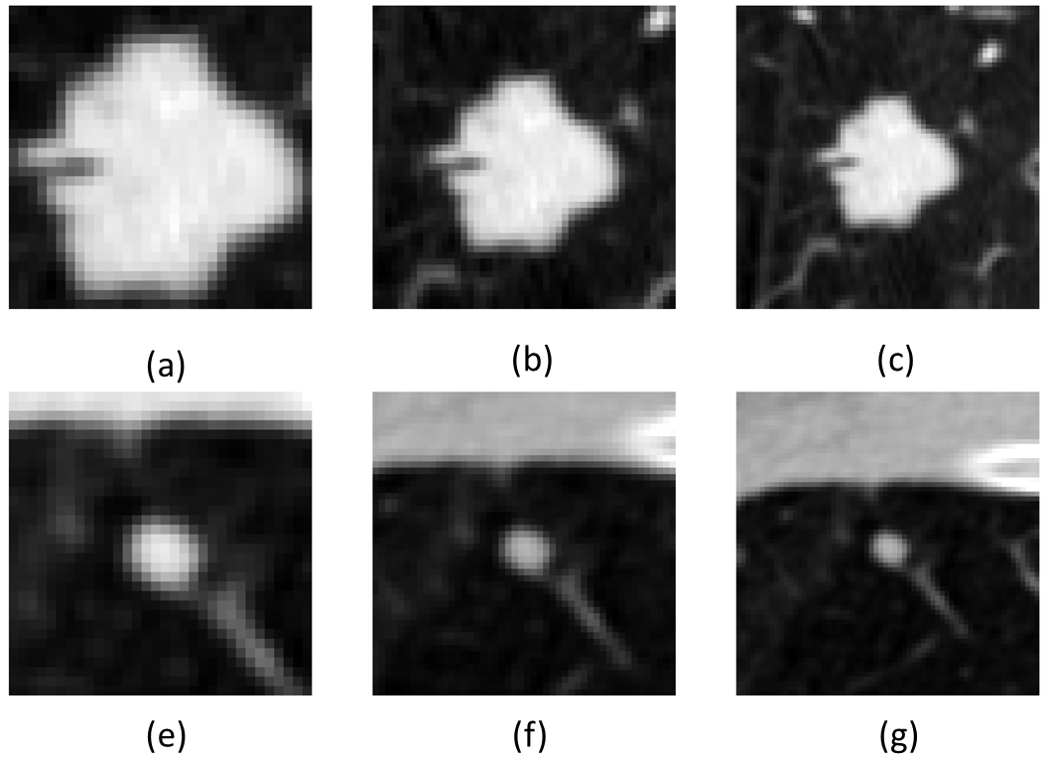}
  \caption{Visual comparison of mid-slices from three cubic patch sizes—32×32×32, 48×48×48, and 64×64×64 voxels—centered on the same lung nodule, where (a),(e) correspond to $32^3$, (b),(f) to $48^3$, and (c),(g) to $64^3$ patches.
}
  \label{PS}
\end{figure}

After evaluating different architectures, we further performed an additional experiment to determine the optimal input patch size when training the proposed Baseline 3D CNN. Since lung nodules vary in size and appearance, the choice of patch dimension can influence classification performance. To investigate this, we extracted three cubic patches of dimensions 32×32×32, 48×48×48, and 64×64×64 voxels centered at each candidate location and trained the Baseline 3D CNN independently on each patch configuration. Table~\ref{tab:patch_compare} reports the corresponding performance metrics. As illustrated in the results, the 32×32×32 patch achieves the highest accuracy, sensitivity, and precision. These findings indicate that smaller patches preserve the most discriminative features for nodule classification. A visual comparison of the mid-slices from the three patch sizes for the same nodule is provided in Fig.\ref{PS}, highlighting the differences in spatial detail and contextual coverage.

\begin{table}[ht]
\centering
\small
\caption{\textbf{Performance of the Baseline 3D CNN Across Different Patch Sizes} \label{tab:patch_compare}}
\setlength{\tabcolsep}{6pt}
\begin{tabular}{p{70pt}|p{20pt}|p{20pt}|p{20pt}|p{20pt}}
\hline
\textbf{Patch Size} & \textbf{\boldmath$\%A_{CC}$} & \textbf{\boldmath$\%S_{EN}$} & \textbf{\boldmath$\%P_{RE}$} & \textbf{\%AUC} \\
\hline
64$\times$64$\times$64 & 84.82 & 84.93 & 85.32 & \textbf{94.10} \\
48$\times$48$\times$48 & 86.61 & 86.52 & 86.93 & 88.42 \\
32$\times$32$\times$32 & \textbf{87.50} & \textbf{88.00} & \textbf{89.00} & 93.63 \\
\hline
\end{tabular}
\end{table}

\subsubsection{Impact of Semi-Supervised Learning}
We employ a semi-supervised learning approach to enhance the robustness of the backbone 3D CNN and improve its classification accuracy. In this semi-supervised technique, we incorporated both labeled and unlabeled data. By leveraging the unlabeled data, the model discerns from a more diverse and \mbox{extensive} training set, which can reduce overfitting and improve generalization to new data. The semi-supervised approach demonstrates its effectiveness through notable improvement in classification accuracy. 

Following the implementation of semi-supervised learning, the accuracy of the fixed test of LUNA16 increases from 87.50\% to 90.18\%. We expand the training dataset significantly, from 355 to 961, contributing to a more diverse, robust training process. In addition to accuracy, other key \mbox{performance} metrics such as sensitivity and specificity also see improvements. These improvements suggest that the model becomes more accurate and better at identifying true positives (increased sensitivity) and reducing false positives (increased specificity). Several key factors drive the enhanced performance, including increased data diversity, mitigation of overfitting, improved feature representation, and increased model robustness. A comparison of test metrices before and after applying semi-supervised technique is given in Table~\ref{tab:semi_supervised}.

\begin{table}[h]
\renewcommand{\arraystretch}{1}
\centering
\caption{\textbf{Performance Comparison of Baseline Models with Fully-Supervised and Semi-Supervised Learning} \label{tab:semi_supervised}} 
%\small
\setlength{\tabcolsep}{6pt}  % reduced from 12pt to allow narrower fit
\begin{tabular}{p{95pt} |p{25pt}| p{25pt}| p{25pt}| p{25pt}}  
\hline

\textbf{\centering Model} & \textbf{\boldmath$\%A_{CC}$} & \textbf{\boldmath$\%P_{RE}$} & \textbf{\boldmath$\%S_{EN}$} & \textbf{\%AUC} \\
\hline
\makecell[l]{Baseline\\(Fully-Supervised)} & 87.50 & 89.00 & 88.00 & \textbf{93.63} \\
\makecell[l]{Baseline\\(Semi-Supervised)}  & \textbf{90.18}  & \textbf{90.00}  & \textbf{90.00}  & 93.09 \\

\hline
\end{tabular}
\end{table}

\subsubsection{Imapact of the Multibranch Model}
To emphasize the focus on the HU-based intensity range, the multi-branch model is implemented. In this approach, the input data is segmented based on varying intensity ranges, and each branch processes data through a separate model pathway. The branching strategy facilitates better feature extraction from the input images, leading to superior performance compared to backbone 3D CNN. We explore various branching configurations, including scenarios with and without the inclusion of the original input, and making the intensity range learnable or non-learnable.
\\

% \textbf{\textit{Non-Learnable Model:}}~%
{\bfseries\itshape Non-Learnable Model:}

When intensity ranges are non-learnable, the model \mbox{performs} poorly than backbone 3D CNN, as illustrated in the Table~\ref{tab:non-learnable}. The fixed intensity thresholds fail to adapt to the specific nuances of the data and cause a decrease in performance, resulting in a less effective representation of critical features needed for accurate classification. The nonadaptive approach of the non-learnable method leads to the loss of important data and a lack of generalization. By not allowing flexibility, the model extracts feature inefficiently and results in lower performance, as evident in Table~\ref{tab:non-learnable}.

\begin{table}[h]
\centering
\caption{\textbf{Performance Comparison of Multi-branch LCC-Net Configurations Using Non-learnable Intensity Intervals} \label{tab:non-learnable}} 
\setlength{\tabcolsep}{7pt}  % consistent with LUNA16 table
\small 
\begin{tabular}{p{90pt} |p{20pt}| p{20pt}| p{20pt}| p{20pt}}
\hline
\textbf{\centering Model Configuration} & \textbf{\boldmath$\%A_{CC}$} & \textbf{\boldmath$\%P_{RE}$} & \textbf{\boldmath$\%S_{EN}$} & \textbf{\%AUC} \\
\hline
11 Branches & 85.71 & 86.49 & 85.71 & 93.27 \\
6 Branches  & 84.82 & 85.12 & 84.82 & 89.19 \\
3 Branches  & 86.60 & 87.22 & 86.61 & 92.06 \\
\hline
\end{tabular}
\end{table}

{\bfseries\itshape Proposed LMLCC-Net:}
% \textbf{\textit{Proposed LMLCC-Net:}}~%
Making the filter range learnable increases the model's adaptability by allowing it to dynamically adjust during training. Unlike explicitly defined ranges, a learnable filter range allows the model to determine the optimal window length based on the data it encounters and extract important features. We tune the range of each branch alongside the model weights, which is a key innovation of our work.

A comparative study of different parameter combinations revealed that the LMLCC-Net with two and three branches consistently outperformed the other variations. The results of these \mbox{comparative} experiments are summarized in Table~\ref{tab:learnable}, highlighting the impact of different branching techniques on model \mbox{performance}. 
\begin{table}[ht]
\centering
\scriptsize

    \caption{\textbf{Performance Comparison of LMLCC-Net Configurations Using Learnable Intensity Intervals.}}
\label{tab:learnable}

    \renewcommand{\arraystretch}{1} 
    \setlength{\tabcolsep}{5pt} 
    \begin{tabular}{c|c|c|c|c|c|c}
        \hline
        \multicolumn{3}{c|}{\textbf{Parameters}}  & \multirow{2}{*}{\textbf{\boldmath$A_{CC}$ \%}} &
        \multirow{2}{*}{\textbf{\boldmath$S_{PE}$ \%}} & \multirow{2}{*}{\textbf{\boldmath$S_{EN}$ \%}} & \multirow{2}{*}{\textbf{AUC\%}} \\
        \cline{1-3}
        \shortstack{\textbf{Number} \\ \textbf{of} \\ \textbf{Branches}} & \shortstack{\textbf{Including} \\ \textbf{Original} \\ \textbf{Input}} & \textbf{Initializer} & & & &  \\
        \hline
        \multirow{3}{*}{2} & X & Random  & 89.29 & 89.31 & 89.31 & 93.46 \\
        & X & Constant & 90.18 & 90.74 & 90.28 & 93.58 \\
        & \checkmark & Constant & 87.50 & 87.66 & 87.56 & 90.71 \\
        \hline
        \multirow{4}{*}{3} & X & Constant & \textbf{91.96} & \textbf{92.24} & \textbf{92.94} & \textbf{94.07} \\
    & X & Random & 88.34 & 86.67 & 85.81 & 90.11 \\ 
    & \checkmark & Constant & 83.04 & 83.52 & 83.14 & 86.92 \\
        \hline
    \end{tabular}
\end{table}

\begin{figure*}[t]
\centering
\includegraphics[width=0.9\textwidth]{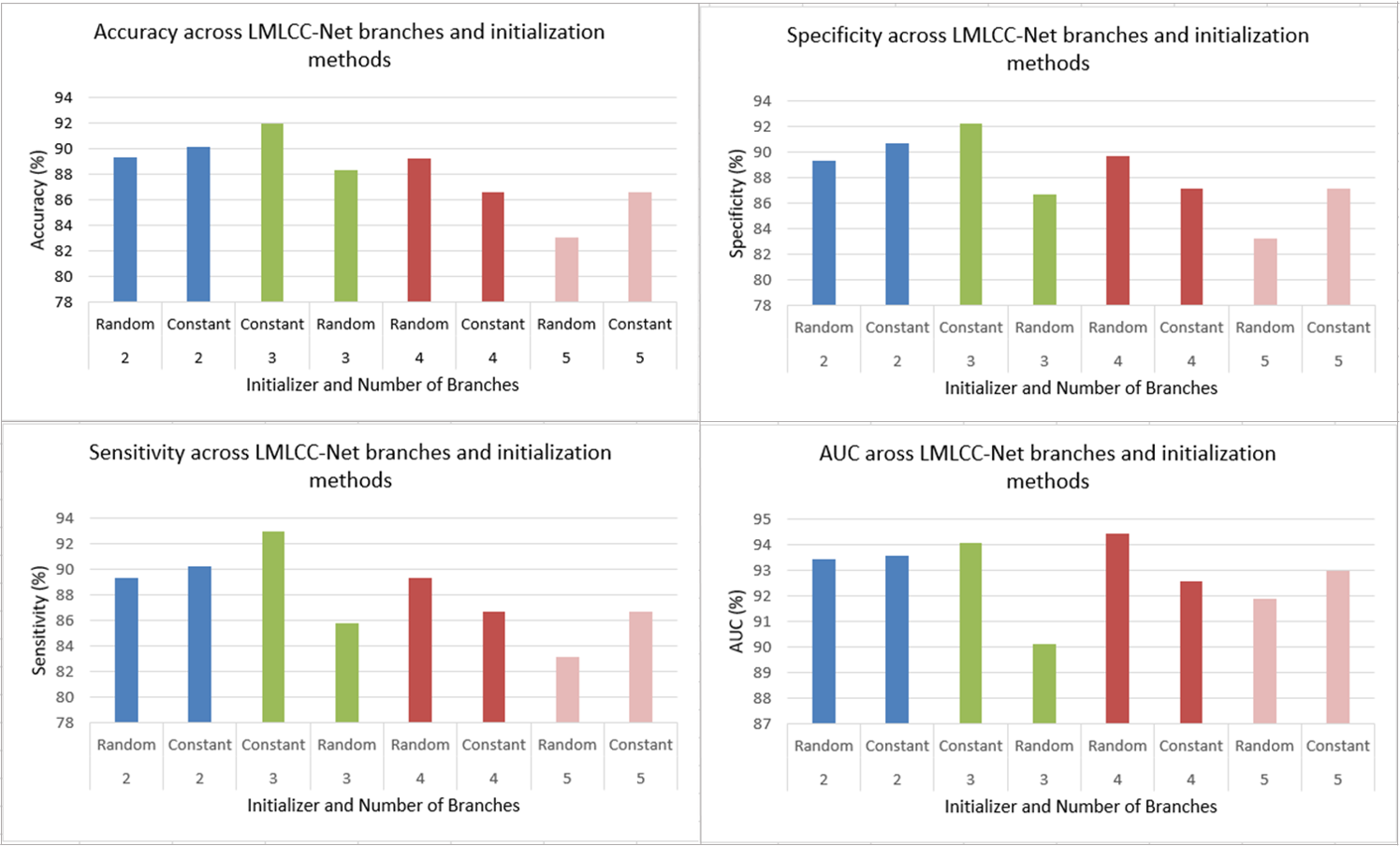}
\caption{Performance comparison of LMLCC-Net across different numbers of branches and initialization methods for without original input. Each bar chart presents accuracy, specificity, sensitivity, and AUC for random and constant initializers with 2–5 branches.}
\label{branch_analysis}
\end{figure*}

Fig. \ref{branch_analysis} further visualizes how varying the number of branches and initialization methods (random vs. constant) affects model performance across four key metrics—accuracy, specificity, sensitivity, and AUC. Each subplot corresponds to one performance metric, clearly illustrating that configurations with two or three branches achieve the best balance across all metrics, while performance begins to decline with four or more branches.  Notably, models initialized with constant intensity filters outperform their randomly initialized counterparts.

Table~\ref{tab:learnable} shows that the three-branch configuration demonstrates the most consistent and superior results, achieving the highest accuracy, specificity, and sensitivity. The constantly initialized approach outperforms the randomly initialized approach. One possible reason is that the randomly initialized model may not yet be able to automatically learn the effective intensity ranges as resources are limited. As a result, the inherent randomness in model initialization can lead to abrupt changes in behavior, causing the model to overlook critical details in important areas and exhibit instability during training.

`

In contrast, using a constant initializer, which accepts a fixed interval and optimizes it within a defined constraint during training, ensures that the interval remains within a practical and stable range. This constraint is critical for maintaining stability during training. Using 2 or 3 intensity ranges seems optimal as it provides the right balance of data granularity for our model. When we use 4, 5, or 6 HU intensity filters, the model struggles to distinguish between the branches due to excessive branching, leading to diminished performance. Furthermore, incorporating the original input alongside HU-filtered branches was found to introduce redundancy and reduce discriminative learning, whereas excluding it yielded more stable and higher performance across all evaluation metrics.

\begin{figure}[!h] 
  \includegraphics[width=3.55 in,height=3 in,clip,keepaspectratio]{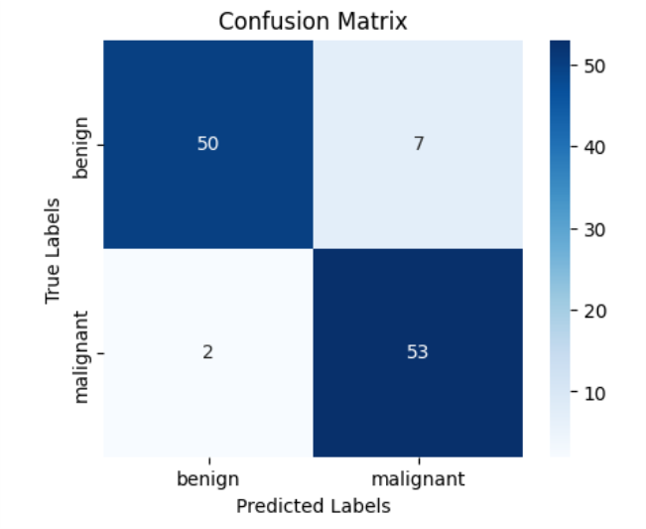}
  \caption{Confusion matrix for the three-branch (constant initializer and without original input) LMLCC-Net configuration.}
  \label{CM}
\end{figure}

To further validate the performance of the best-performing configuration, a confusion matrix is generated, as illustrated in Fig. \ref{CM}. It visualizes the distribution of predicted versus true labels, providing a clear understanding of the model’s classification reliability. The matrix indicates that the model accurately identified 50 benign and 53 malignant nodules, misclassifying only a few cases (7 benign as malignant and 2 malignant as benign). This demonstrates the model’s strong balance between sensitivity and specificity and its robustness in differentiating benign from malignant nodules—an essential requirement for clinical deployment.

\begin{figure}[!h] 
  \includegraphics[width=3.55 in,height=3 in,clip,keepaspectratio]{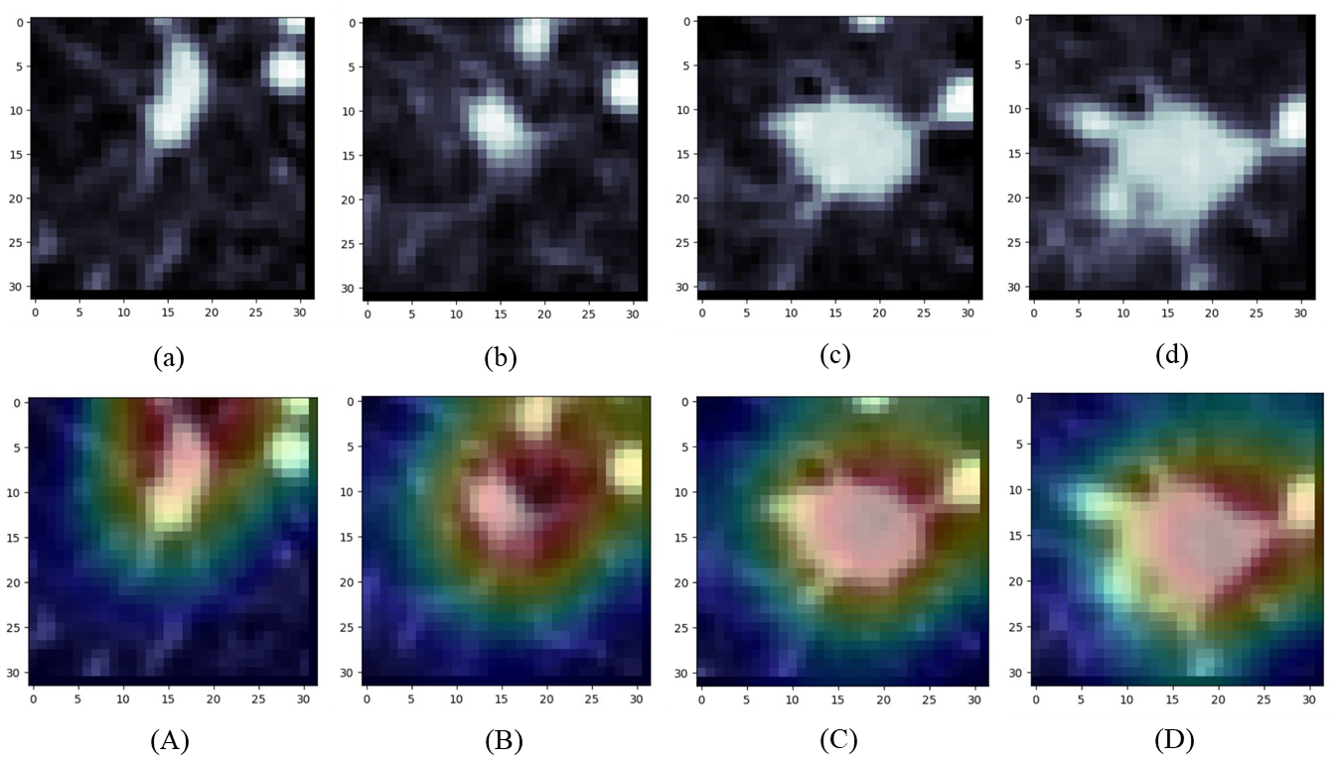}
  \caption{Example lung nodule slices (a-d) and corresponding Grad-CAM heatmaps (A-D) produced by the proposed LMLCC-Net.}
  \label{grad}
\end{figure}

To further interpret the learned features and validate that the proposed LMLCC-Net focuses on diagnostically relevant regions, we employed Gradient-weighted Class Activation Mapping (Grad-CAM) on several test nodules as shown in Fig.~\ref{grad}. These heatmaps highlight the regions contributing most to the malignancy prediction, allowing qualitative verification of the model’s attention on the nodule regions rather than irrelevant background artifacts.

\subsubsection{Benchmarking Against Prior Works}
To benchmark our proposed model against published classification models for pulmonary nodules, we train LMLCC-Net on the LUNA16 dataset. We evaluate performance using the metrics \( A_{CC} \), \( S_{PE} \), and \( S_{EN} \). A comparative analysis of the literature presented in Table~\ref{tab:literature} reveals that most existing models demonstrate limited effectiveness in nodule classification. In contrast, LMLCC-Net achieves superior results across all of the performance metrics, \( A_{CC} \), \( S_{PE} \), and \( S_{EN} \).

\begin{table}[h!]
\centering
\caption{\textbf{Comparison of classification performance reported across various methods for pulmonary nodule detection on the LUNA16 dataset.}}
\label{tab:literature}
\setlength{\tabcolsep}{6pt}

\resizebox{\linewidth}{!}{

\begin{tabular}{p{65pt}|>{\centering\arraybackslash}m{25pt}|p{25pt}|p{25pt}|p{45pt}}
\hline
\textbf{Method$^{*}$} & \textbf{\boldmath$\%A_{CC}$} & \textbf{\boldmath$\%S_{PE}$} &
\textbf{\boldmath$\%S_{EN}$} & \textbf{Model Input} \\
\hline
Shen et al.~\cite{shen2017multi}         & 87.14 & 77.00 & 93.00   & 64$^3$ \\
Xie et al.~\cite{xie2018knowledge}       & 91.60 & 86.52 & 94.00  & 64$^3$ \\
Shen et al.~\cite{shen2019interpretable} & 84.20 & 70.50 & 88.90  & 52$^3$ \\
Parnian et al.~\cite{afshar2021mixcaps}  & 90.70 & 89.50 & 89.50   & -- \\
Liu et al.~\cite{liu2019multi}           & 90.10 & 88.50 & 90.50   & 224$^2$ \\
Nasrullah et al.~\cite{nasrullah2019automated} & 88.79 & 89.83 & 93.97 & 36$^3$ \\
Lijing et al.~\cite{sun2024nodule}       & 90.60 & 92.60 & 92.81  & 32$^3$ \\
Gupta et al.~\cite{gupta2024}            & 91.30 & 92.10 & 90.00 & 64$^2\times$32 \\
Liu et al.~\cite{liu2020multi}           & 90.20 & 92.90 & 83.70  & 16$^3$, 32$^3$, 64$^3$ \\
\hline
\textbf{LMLCC-Net}                       & \textbf{91.96} & \textbf{92.24} & \textbf{92.94}   & \textbf{32$^3$} \\
\hline
\end{tabular}%
}

\raggedright
\scriptsize{$^{*}$Results are reported for reference only. Different methods use different train–test splits, so direct comparison is not exact.}
\end{table}

In a clinical setting, providers must achieve a balance between $S_{PE}$ and $S_{EN}$. While an increased $S_{EN}$ improves the detection rate of malignant nodules, a lower $S_{PE}$ can lead to a higher number of false positives, resulting in unnecessary follow-up procedures and patient anxiety. We maintain a balanced dataset with nearly equal numbers of malignant and benign samples at the beginning, thus resulting in a balanced value of $S_{PE}$ and $S_{EN}$. Although CLIP-based models for lung diagnosis \cite{sun2024nodule} have made significant
progress, they still face key limitations. First, these models rely on text attributes annotated by radiologists, whose analysis of CT images is inherently subjective and prone to errors. Moreover, since these annotations are only utilized during training, they do not offer direct applicability during the inference
phase. Second, the vision encoder backbone used in Nodule-CLIP \cite{sun2024nodule} is based on ResNet with randomly initialized weights. This approach contradicts the fundamental principle of contrastive pretraining, as randomized initialization disregards prior knowledge that could enhance feature extraction. 

LMLCC-Net introduces a noble approach by using the Hounsfield Unit as a training parameter. The proposed model offers an improved balance between sensitivity and specificity. This balance is essential for practical use and to improve the reliability and clinical utility of the model. 

\section{Conclusion}
\label{sec:conclusion}
In this study, we introduce a novel approach to address the challenge of ambiguous labels within the LUNA16 dataset, a subset of LIDC-IDRI, by implementing a semi-supervised labeling method. This approach significantly improves the accuracy and reliability of the dataset, providing a more solid foundation for training machine learning models. Building on this foundation, we propose LMLCC-Net to enhance the prediction of lung nodule malignancy on CT images. It processes normalized CT scans by dividing the input into branches that learn and focus on specific intensity ranges corresponding to the CT Hounsfield Units. This design allows the model to capture richer spatial information and extract more representative features, ultimately improving its ability to differentiate between benign and malignant nodules. Our contributions demonstrate the effectiveness of combining innovative labeling techniques with advanced neural network architectures. The LMLCC-Net offers a promising tool for more accurate and reliable lung nodule classification, which could have significant implications for early cancer detection and patient outcomes.

\begin{IEEEbiography}[{\includegraphics[width=1in,height=1.25in,clip,keepaspectratio]{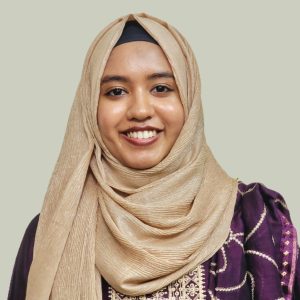}}]{Tasnia Binte Mamun}  has recently pursued the B.Sc. degree in Biomedical Engineering at Bangladesh University of Engineering and Technology (BUET), Bangladesh. She is currently working as a research assistant in the mHealth Lab, BUET. Her interest lies on Biomedical signal processing, AI and Machine Learning in Healthcare, Medical Image Processing, MRI.
\end{IEEEbiography}

\begin{IEEEbiography}[{\includegraphics[width=1in,height=1.25in,clip,keepaspectratio]{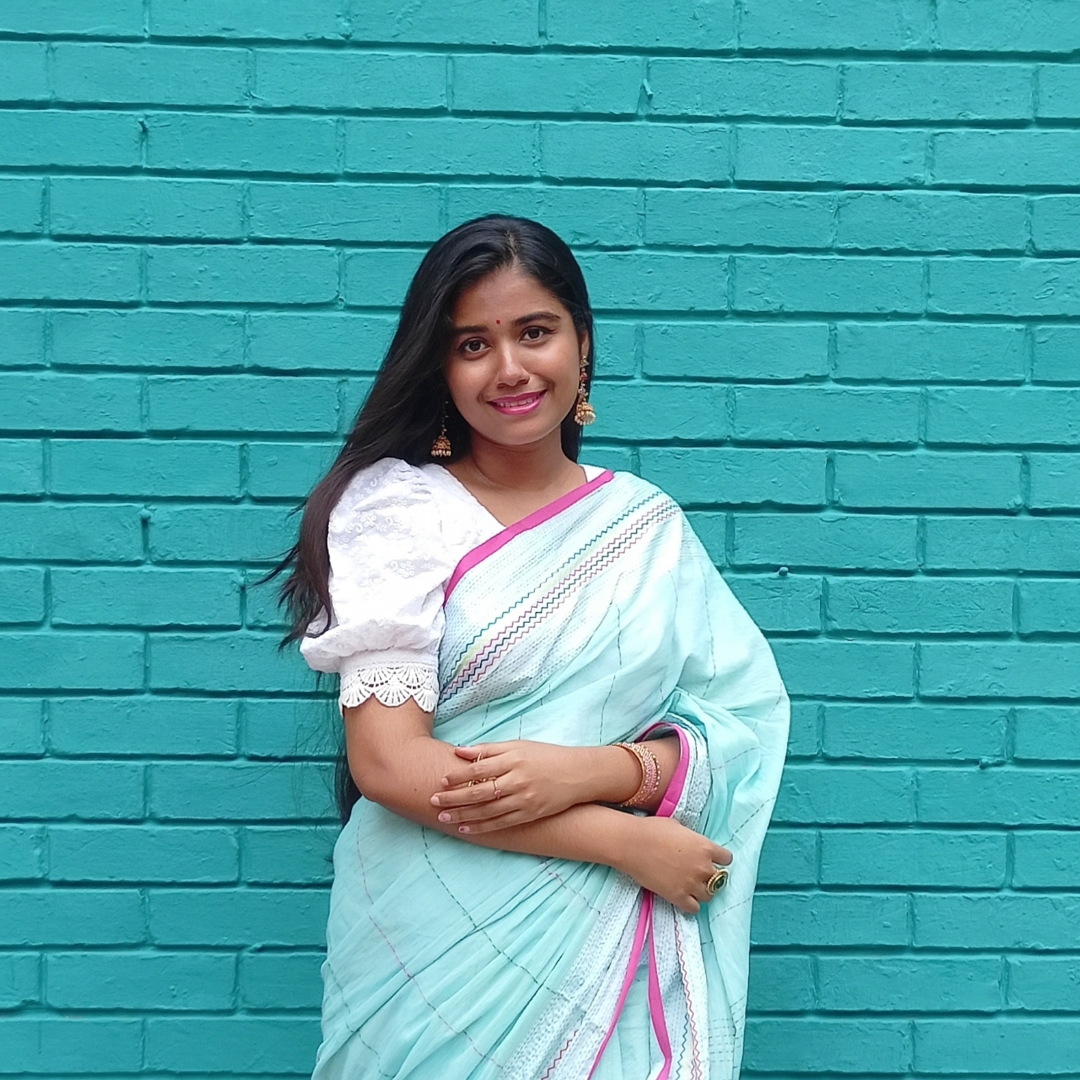}}]{Adhora Madhuri} received the B.Sc. degree in Biomedical Engineering from Bangladesh University of Engineering and Technology (BUET), Bangladesh, in 2024. Her research includes computer vision and machine learning for use in low-resource settings. Her other interest lies in the world of material sciences and using engineering knowledge to create better healthcare solutions.
\end{IEEEbiography}

\begin{IEEEbiography}[{\includegraphics[width=1in,height=1.25in,clip,keepaspectratio]{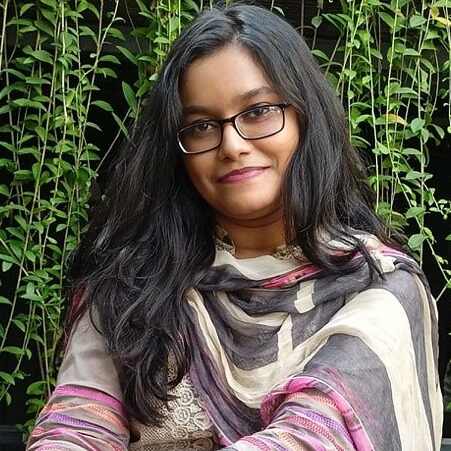}}]{Nusaiba Sobir}  received the B.Sc. degree in Biomedical Engineering from Bangladesh University of Engineering and Technology (BUET), Bangladesh, in 2024. Her research focuses on biomedical signal processing, image processing, and machine learning applications in low-resource settings. She is also interested in material science and leveraging engineering principles to develop improved healthcare solutions.
\end{IEEEbiography}

\begin{IEEEbiography}[{\includegraphics[width=1in,height=1.25in,clip,keepaspectratio]{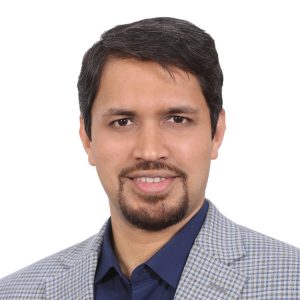}}]{Dr. Taufiq Hasan} (Senior member, IEEE) completed his BSc. and MSc. in Electrical and Electronic Engineering (EEE) from Bangladesh University of Engineering and Technology (BUET). He obtained his PhD. in Electrical En gineering from The University of Texas at Dallas, where he was a member of the Center of Robust Speech Systems(CRSS). He worked as a Research Scientist at Robert Bosch Research and Technology Center, Palo Alto, CA. Dr. Hasan’s primary affiliation is with the Department of Biomedical Engineering at BUET as a Professor,where he leads the mHealth research group. His research interests are biomedical signal/image analysis, and medical device design.
\end{IEEEbiography}

\EOD


\begin{thebibliography}{00}

\bibitem{bray2024global}
Freddie Bray, Mathieu Laversanne, Hyuna Sung, Jacques Ferlay, Rebecca L. Siegel, Isabelle Soerjomataram, Ahmedin Jemal,\textit{Global cancer statistics 2022: GLOBOCAN estimates of incidence and mortality worldwide for 36 cancers in 185 countries}, CA Cancer J. Clin., vol. 74, no. 3, pp. 229--263, 2024.

\bibitem{parkin2000lung}
D. Maxwell Parkin, Sue M. Moss,\textit{Lung cancer screening: improved survival but no reduction in deaths—the role of “overdiagnosis”},Cancer, vol. 89, no. S11, pp. 2369--2376, 2000.

\bibitem{rubin2015lung}
Geoffrey D. Rubin, \textit{Lung nodule and cancer detection in computed tomography screening}, J. Thorac. Imaging, vol. 30, no. 2, pp. 130--138, 2015.

\bibitem{lancaster2022low}
Harriet L. Lancaster, Marjolein A. Heuvelmans, Matthijs Oudkerk, \textit{Low-dose computed tomography lung cancer screening: Clinical evidence and implementation research}, J. Intern. Med., vol. 292, no. 1, pp. 68--80, 2022.

\bibitem{loverdos2019lung}
Konstantinos Loverdos, Andreas Fotiadis, Chrysoula Kontogianni, Marianthi Iliopoulou, Mina Gaga, \textit{Lung nodules: a comprehensive review on current approach and management}, Ann. Thorac. Med., vol. 14, no. 4, pp. 226--238, 2019.

\bibitem{ojala2002multiresolution}
Timo Ojala, Matti Pietikainen, Topi Maenpaa, \textit{Multiresolution gray-scale and rotation invariant texture classification with local binary patterns}, IEEE Trans. Pattern Anal. Mach. Intell., vol. 24, no. 7, pp. 971--987, 2002.

\bibitem{chen2018radiomic}
Chia-Hung Chen, Chih-Kun Chang, Chih-Yen Tu, Wei-Chih Liao, Bing-Ru Wu, Kuei-Ting Chou, Yu-Rou Chiou, Shih-Neng Yang, Geoffrey Zhang, Tzung-Chi Huang, \textit{Radiomic features analysis in computed tomography images of lung nodule classification}, PLOS ONE, vol. 13, no. 2, e0192002, 2018.

\bibitem{thawani2018radiomics}
Rajat Thawani, Michael McLane, Niha Beig, Soumya Ghose, Prateek Prasanna, Vamsidhar Velcheti, Anant Madabhushi, \textit{Radiomics and radiogenomics in lung cancer: a review for the clinician}, Lung Cancer, vol. 115, pp. 34--41, 2018.

\bibitem{wang2023diagnostic}
Hongfeng Wang, Hai Zhu, Lihua Ding, Kaili Yang, \textit{A diagnostic classification of lung nodules using multiple-scale residual network}, Sci. Rep., vol. 13, no. 1, p. 11322, 2023.

\bibitem{xie2018knowledge}
Yutong Xie, Yong Xia, Jianpeng Zhang, Yang Song, Dagan Feng, Michael Fulham, Weidong Cai, \textit{Knowledge-based collaborative deep learning for benign-malignant lung nodule classification on chest CT}, IEEE Trans. Med. Imaging, vol. 38, no. 4, pp. 991--1004, 2018.

\bibitem{asuntha2020deep}
A. Asuntha, Andy Srinivasan, \textit{Deep learning for lung cancer detection and classification}, Multimed. Tools Appl., vol. 79, no. 11, pp. 7731--7762, 2020.

\bibitem{shen2019interpretable}
Shiwen Shen, Simon X. Han, Denise R. Aberle, Alex A. Bui, William Hsu, \textit{An interpretable deep hierarchical semantic convolutional neural network for lung nodule malignancy classification}, Expert Syst. Appl., vol. 128, pp. 84--95, 2019.

\bibitem{sori2021dfd}
Worku J. Sori, Jiang Feng, Arero W. Godana, Shaohui Liu, Demissie J. Gelmecha, \textit{DFD-Net: lung cancer detection from denoised CT scan image using deep learning}, Front. Comput. Sci., vol. 15, pp. 1--13, 2021.

\bibitem{xu2020mscs}
Xiuyuan Xu, Chengdi Wang, Jixiang Guo, Yuncui Gan, Jianyong Wang, Hongli Bai, Lei Zhang, Weimin Li, Zhang Yi, \textit{MSCS-DeepLN: Evaluating lung nodule malignancy using multi-scale cost-sensitive neural networks}, Med. Image Anal., vol. 65, p. 101772, 2020.

\bibitem{naseer2023lung}
Iftikhar Naseer, Sheeraz Akram, Tehreem Masood, Muhammad Rashid, Arfan Jaffar, \textit{Lung cancer classification using modified U-Net based lobe segmentation and nodule detection}, IEEE Access, vol. 11, pp. 60279--60291, 2023.

\bibitem{li2019predicting}
Shulong Li, Panpan Xu, Bin Li, Liyuan Chen, Zhiguo Zhou, Hongxia Hao, Yingying Duan, Michael Folkert, Jianhua Ma, Shiying Huang, et al., \textit{Predicting lung nodule malignancies by combining deep convolutional neural network and handcrafted features}, Phys. Med. Biol., vol. 64, no. 17, p. 175012, 2019.

\bibitem{zhang2019lung}
Qianqian Zhang, Haifeng Wang, Sang Won Yoon, Daehan Won, Krishnaswami Srihari, \textit{Lung nodule diagnosis on 3D computed tomography images using deep convolutional neural networks}, Procedia Manuf., vol. 39, pp. 363--370, 2019.

\bibitem{tumorlist}
National Cancer Institute (U.S.),\textit{Tumor List}, \url{https://training.seer.cancer.gov/disease/categories/tumors.html}, Accessed 8/15/2024.

\bibitem{Benigntumors}
Medline Plus,\textit{Benign Tumors}, \url{https://medlineplus.gov/benigntumors.html}, Accessed 8/15/2024.

\bibitem{patel2020benign}
Aisha Patel,\textit{Benign vs malignant tumors}, JAMA Oncol., vol. 6, no. 9, p. 1488, 2020.

\bibitem{ohshika2021distinction}
Shusa Ohshika, Tatsuro Saruga, Tetsuya Ogawa, Hiroya Ono, Yasuyuki Ishibashi, \textit{Distinction between benign and malignant soft tissue tumors based on an ultrasonographic evaluation of vascularity and elasticity},Oncol. Lett., vol. 21, no. 4, p. 1, 2021.

\bibitem{setio2016pulmonary}
Arnaud Arindra Adiyoso Setio, Francesco Ciompi, Geert Litjens, Paul Gerke, Colin Jacobs, Sarah J. Van Riel, Mathilde Marie Winkler Wille, Matiullah Naqibullah, Clara I. Sánchez, Bram Van Ginneken,\textit{Pulmonary nodule detection in CT images: false positive reduction using multi-view convolutional networks}, IEEE Trans. Med. Imaging, vol. 35, no. 5, pp. 1160--1170, 2016.

\bibitem{messay2015segmentation}
Temesguen Messay, Russell C. Hardie, Timothy R. Tuinstra, \textit{Segmentation of pulmonary nodules in computed tomography using a regression neural network approach and its application to the lung image database consortium and image database resource initiative dataset}, Med. Image Anal., vol. 22, no. 1, pp. 48--62, 2015.

\bibitem{guo20233d}
Zhitao Guo, Jikai Yang, Linlin Zhao, Jinli Yuan, and Hengyong Yu, \textit{3D SAACNet with GBM for the classification of benign and malignant lung nodules}, Computers in Biology and Medicine, Elsevier, Vol. 153, 2023, pp. 106532.

\bibitem{kumar2021classification}
Vinod Kumar and Brijesh Bakariya, \textit{Classification of malignant lung cancer using deep learning}, Journal of Medical Engineering \& Technology, Taylor \& Francis, Vol. 45, No. 2, 2021, pp. 85--93.

\bibitem{dai2018incorporating}
Yaojun Dai, Shiju Yan, Bin Zheng, and Chengli Song, \textit{Incorporating automatically learned pulmonary nodule attributes into a convolutional neural network to improve accuracy of benign-malignant nodule classification}, Physics in Medicine \& Biology, IOP Publishing, Vol. 63, No. 24, 2018, pp. 245004.

\bibitem{kuang2020unsupervised}
Yan Kuang, Tian Lan, Xueqiao Peng, Gati Elvis Selasi, Qiao Liu, and Junyi Zhang, \textit{Unsupervised multi-discriminator generative adversarial network for lung nodule malignancy classification}, IEEE Access, Vol. 8, 2020, pp. 77725--77734.

\bibitem{joshi2021lung}
Aniket Joshi, Jayanthi Sivaswamy, and Gopal Datt Joshi, \textit{Lung nodule malignancy classification with weakly supervised explanation generation}, Journal of Medical Imaging, SPIE, Vol. 8, No. 4, 2021, pp. 044502.

\bibitem{causey2018highly}
Jason L. Causey et al., \textit{Highly accurate model for prediction of lung nodule malignancy with CT scans}, Scientific Reports, Nature Publishing Group, Vol. 8, 2018, pp. 9286.

\bibitem{apostolopoulos2021classification}
Ioannis D. Apostolopoulos, Nikolaos D. Papathanasiou, and George S. Panayiotakis,
\textit{Classification of lung nodule malignancy in computed tomography imaging utilising generative adversarial networks and semi-supervised transfer learning}, Biocybernetics and Biomedical Engineering, Elsevier, Vol. 41, No. 4, 2021, pp. 1243--1257.

\bibitem{qin2021relationship}
Yulei Qin et al., \textit{Relationship between pulmonary nodule malignancy and surrounding pleurae, airways and vessels: a quantitative study using the public LIDC-IDRI dataset},
arXiv preprint arXiv:2106.12991, 2021.

\bibitem{zhang2022re}
Hanxiao Zhang et al.,
\textit{Re-thinking and re-labeling LIDC-IDRI for robust pulmonary cancer prediction},
In: Workshop on Medical Image Learning with Limited and Noisy Data, Springer, 2022, pp. 42--51.

\bibitem{tang2021classification}
Siyuan Tang et al., \textit{Classification of benign and malignant pulmonary nodules based on the multiresolution 3D DPSECN model and semisupervised clustering}, IEEE Access, Vol. 9, 2021, pp. 43397--43410.

\bibitem{shen2017multi}
Wei Shen et al., \textit{Multi-crop convolutional neural networks for lung nodule malignancy suspiciousness classification},
Pattern Recognition, Elsevier, Vol. 61, 2017, pp. 663--673.

\bibitem{liu2020multi}
Hong Liu et al., \textit{Multi-model ensemble learning architecture based on 3D CNN for lung nodule malignancy suspiciousness classification}, Journal of Digital Imaging, Springer, Vol. 33, 2020, pp. 1242--1256.

\bibitem{afshar2021mixcaps}
Parnian Afshar et al., \textit{MIXCAPS: A capsule network-based mixture of experts for lung nodule malignancy prediction}, Pattern Recognition, Elsevier, Vol. 116, 2021, pp. 107942.

\bibitem{lidc}
LIDC-IDRI: A Complete Reference Database of Lung Nodules on CT Scans. \emph{The Cancer Imaging Archive}. Available at: \url{https://www.cancerimagingarchive.net/collection/lidc-idri/}

\bibitem{luna}
\textit{LUNA16 Grand Challenge Dataset},
Grand Challenge, Available at:\url{https://luna16.grand-challenge.org/Data/}

\bibitem{adam}
Adam Optimizer.\emph{Cornell University}. Available at: \url{https://optimization.cbe.cornell.edu/index.php?title=Adam}

\bibitem{HU}
Hounsfield Unit. \emph{Radiopaedia}. Available at: \url{https://radiopaedia.org/articles/hounsfield-unit}



\bibitem{loss}
Analytics Vidhya. \emph{Understanding Loss Function in Deep Learning}, 2022.  
Available at: \url{https://www.analyticsvidhya.com/blog/2022/06/understanding-loss-function-in-deep-learning/}.



\bibitem{teague2006}
Shawn D. Teague and Dewey J. Conces. \emph{Diagnosis of Lung Cancer: Perspective of a Pulmonary Radiologist}. PET Clinics, Vol. 1, No. 4, pp. 289–300, 2006. Available at: \url{https://doi.org/10.1016/j.cpet.2006.09.004}


\bibitem{kim2011}
Anthony W. Kim and David T. Cooke, 
\textit{Additional pulmonary nodules in the patient with lung cancer: controversies and challenges}, Clin. Chest Med., Vol. 32, No. 4, 2011. \url{https://pubmed.ncbi.nlm.nih.gov/22054888/}

\bibitem{nasrullah2019automated}
Nasrullah Nasrullah et al., 
\textit{Automated lung nodule detection and classification using deep learning combined with multiple strategies}, Sensors, MDPI, Vol. 19, No. 17, 2019, p. 3722.


\bibitem{chandrasekar2022}
Thaventhiran Chandrasekar et al., \textit{Lung cancer disease detection using service-oriented architectures and multivariate boosting classifier}, Applied Soft Computing, Vol. 122, 108820, 2022.

\bibitem{ramos2017}
Juan Ramos-González et al., \textit{A CBR framework with gradient boosting based feature selection for lung cancer subtype classification}, Comput. Biol. Med., Vol. 86, pp. 98–106, 2017.

\bibitem{donga2022}
Harsha Vardhan Donga et al., \textit{Effective framework for pulmonary nodule classification from CT images using the modified gradient boosting method}, Applied Sciences, Vol. 12, No. 16, p. 8264, 2022.

\bibitem{ma2020}
Baoshan Ma et al., \textit{Diagnostic classification of cancers using extreme gradient boosting algorithm and multi-omics data}, Comput. Biol. Med., Vol. 121, 103761, 2020.

\bibitem{hamed2023}
Esraa A-R Hamed et al., \textit{An efficient combination of convolutional neural network and LightGBM algorithm for lung cancer histopathology classification}, Diagnostics, Vol. 13, No. 15, p. 2469, 2023.

\bibitem{yu2020}
Daping Yu et al., \textit{Copy number variation in plasma as a tool for lung cancer prediction using Extreme Gradient Boosting (XGBoost) classifier}, Thoracic Cancer, Vol. 11, No. 1, pp. 95–102, 2020.

\bibitem{hara2017}
Kensho Hara, Hirokatsu Kataoka, and Yutaka Satoh, \textit{Learning spatio-temporal features with 3D residual networks for action recognition}, Proc. of the IEEE Int. Conf. on Computer Vision Workshops, pp. 3154–3160, 2017.

\bibitem{zunair2020uniformizing}
Hasib Zunair et al., \textit{Uniformizing techniques to process CT scans with 3D CNNs for tuberculosis prediction}, Proc. of PRIME 2020 (MICCAI Workshop), pp. 156–168, 2020.

\bibitem{sun2024nodule}
Lijing Sun et al., \textit{Nodule-CLIP: Lung nodule classification based on multi-modal contrastive learning}, Comput. Biol. Med., Vol. 175, 108505, 2024.

\bibitem{gupta2024}
Himanshu Gupta, Himanshu Singh, and Anil Kumar, 
\textit{Texture and radiomics inspired data-driven cancerous lung nodules severity classification}, Biomed. Signal Process. Control, Vol. 88, 105543, 2024.

\bibitem{liu2019multi}
Lihao Liu et al., \textit{Multi-task deep model with margin ranking loss for lung nodule analysis}, IEEE Trans. Med. Imaging, IEEE, Vol. 39, No. 3, 2019, pp. 718--728.

\end{thebibliography}
\end{document}